# Strongly enhanced critical current density in MgB$_2$/Fe tapes by stearic acid and stearate doping

Zhaoshun Gao[1], Yanwei Ma[1,*], Xianping Zhang[1], Dongliang Wang[1], Zhengguang Yu[1], K. Watanabe[2], Huan Yang[3], Haihu Wen[3]

[1]Applied Superconductivity Lab., Institute of Electrical Engineering, Chinese Academy of Sciences, P. O. Box 2703, Beijing 100080, China

[2] High Field Laboratory for Superconducting Materials, Institute for Materials Research, Tohoku University, Sendai 980-8577, Japan

[3]Institute of Physics, Chinese Academy of Sciences, Beijing 100080, China

**Abstract:**

Fe-sheathed MgB$_2$ tapes with cheap stearic acid, Mg stearate and Zn stearate doping were prepared through the in situ powder-in-tube method. It is found that $J_c$, $H_{irr}$, and $H_{c2}$ were significantly enhanced by doping. Compared to the pure tapes, J$_c$ for all the doped samples was improved by more than an order of magnitude in a high-field regime. At 4.2 K, the transport $J_c$ reached $2.02 \times 10^4$ A/cm$^2$ at 10 T for the Zn stearate doped tapes and $3.72 \times 10^3$ A/cm$^2$ at 14 T for the stearic acid doped samples, respectively. Moreover, at 20 K, $H_{irr}$ for the Zn stearate doped tape achieved 10 T, which is comparable to that of the commercial NbTi at 4.2K. The improvement of $J_c$-$H$ properties in doped samples can be attributed to the increase of $H_{c2}$ resulting from the incorporation of C atoms into the MgB$_2$ crystal lattice as well as a high density of flux-pinning centers.

[*] Author to whom correspondence should be addressed; E-mail: ywma@mail.iee.ac.cn

## 1. Introduction

MgB$_2$ has been regarded as a possible substitute for the Nb-based conventional superconductors due to its high critical temperature ($T_c$), low material cost and weak-link free behavior. Furthermore, a $T_c$ of 39 K is high enough for use at elevated temperatures of 20 K as the conductor of cryogen-free magnet [1]. Therefore, MgB$_2$ is believed to be a promising candidate for practical applications, especially for MRI magnet in the temperature range of 20-25 K. Unfortunately, the MgB$_2$ still shows lower in-field critical current density $J_c$ values so far because of low upper critical field ($H_{c2}$) and poor flux pinning.

A number of experimental techniques have been attempted to improve the $H_{c2}$ and current carrying ability of MgB$_2$ superconducting materials [2-9]. Among all the methods, doping with carbon or carbon contained compounds seems to be the most effective way to increase the $H_{c2}$ by shortening the mean free length of the electrons, or selective tuning of impurity scattering of the $\pi$ and $\sigma$ bands according to the two-gap model of $H_{c2}$ of Gurevich [10, 11]. Very recently, Dou's group has reported that malic acid doped MgB$_2$ bulks exhibited a positive effect in improving the $J_c$ values through magnetic measurement [12]. However, the effect of carbohydrate doping on $J_c$-$H$ properties of Fe-sheathed MgB$_2$ tapes has not been reported by far. In this work, we used stearic acid and stearate as the dopants to enhance $J_c$-$H$ performance in Fe-sheathed MgB$_2$ tapes. In fact, besides the advantages such as low cost, production of highly reactive C and avoidance of agglomeration for carbohydrate dopants mentioned in Ref [12], there are other two unique merits of stearic acid and stearate doping: First, stearic acid and stearate's soap-like character enables them to reduce the friction between the core and the tube walls during deformation process, it may benefit for the extent of deformation as well as the core homogeneity of PIT tapes. Second, as the representatives of organic acid metallic salt, stearate contains not only carbon but also metal elements. Hence, if we select the proper organic acid metallic salt, carbon-metal co-doped effects might be easily introduced. In this letter, we have synthesized stearic acid and stearate doped MgB$_2$ tapes and succeeded in significantly improving the transport $J_c$-$H$ properties of Fe-sheathed $MgB_2$ tapes in high magnetic fields.

**2. Experimental details**

The $MgB_2$ composite tapes with stearic acid ($C_{18}H_{36}O_2$, 99%), Mg stearate ($C_{36}H_{70}MgO_4$, 98%) and Zn stearate ($C_{36}H_{70}ZnO_4$, 98%) doping were prepared by the in situ PIT method. The added amounts of stearic acid and stearate into the $MgB_2$ samples were 10wt % of total $MgB_2$. Stearic acid and stearate were dissolved in acetone or alcohol. The solution was mixed with an appropriate amount of B powder (amorphous, 99.99%) by ultrasonator with the aim to achieve a more uniform distribution. The mixture was dried at 60 ℃ in a vacuum oven. After drying, the obtained mixture was mixed with an appropriate amount Mg powder(325 mesh, 99.8%). The powder was ground by hand with a mortar and pestle. These mixed powders were packed into Fe tubes, and then cold-rolled into tapes. The final size of the tapes was about 3.8 mm in width and about 0.5 mm in thickness. Pure tapes were also fabricated under the same conditions for comparative study. These tapes were heated at 800 ℃ or 850 ℃ for 1 h, and then followed by a furnace cooling to room temperature. The argon gas was allowed to flow into the furnace during the heat-treatment process to reduce the oxidation of the samples.

The phase identification and crystal structure investigation were carried out using x-ray diffraction (XRD). Microstructure was studied using a transmission electron microscopy (TEM) equipped with an x-ray energy dispersion spectrum (EDS). DC magnetization measurements were performed with a superconducting quantum interference device (SQUID) magnetometer. After peeling away the Fe sheath, resistivity curves were measured with 5 mA transport currents by an Oxford cryogenic system (Maglab-12), the 10% and 90% points on the resistive transition curves were used to define the $H_{c2}$ and $H_{irr}$. The transport current $I_c$ at 4.2 K and its magnetic field dependence were evaluated at the High field laboratory for Superconducting Materials (HFLSM) at Sendai, by a standard four-probe resistive method, with a criterion of 1 $\mu V\ cm^{-1}$, a magnetic field up to 14 T was applied parallel to the tape surface. For each set of doping tape, the Ic measurement was performed for several samples to check reproducibility.

## 3. Results and discussion

Figure 1 shows the XRD patterns for the four kinds of tapes heated at 800℃. These are pure, stearic acid, Mg stearate and Zn stearate doped tapes, respectively. For the pure sample, $MgB_2$ was obtained as a main phase with less amounts of MgO and $MgB_4$. The same was found to be the case for all doped tapes. It is found that the position of both (100) and (110) peaks of doped samples clearly shifts to higher angles, suggesting that the substitution of C in the B site actually occurred. Based on the a-axis lattice parameter changes [13], we could quantitatively estimate the amount of C substituted in B sites, as shown in Table I. The highly actual C substitution levels of 1.94, 1.98 and 1.64 at.% of B at three doped samples indicated that stearic acid and stearate are effective carbon sources for $MgB_2$ doping, which is due to the high reactivity of C released from the decomposition of the dopants. Further, the full width at half-maximum (FWHM) values of the (110) peak for the doped tapes are apparently larger than that of the corresponding peak for the pure one (see Table I). This broadening of the FWHM indicates grain size refinement and lattice distortion of the $MgB_2$, which usually resulted in an enhancement of the flux pinning [14].

Figure 2 shows the temperature dependences of magnetization for the pure and doped samples synthesized at 800 °C. The highest transition temperature ($T_c$ onset = 37.9 K) is observed in the pure $MgB_2$ tapes. For all the doped samples, a small depression in $T_c$ was observed [see Table I]. As reported by Wilke et al. [15], $T_c$ may be used as an indicator of how much carbon is incorporated into the $MgB_2$. Therefore, these results suggest that some amounts of carbon atoms of stearic acid, Mg stearate and Zn stearate were substituted in the B position in our doped samples. It is interesting to note that a smaller decrease of $T_c$ was found in the Zn stearate doped sample compared to other two doped tapes, suggesting that a trace amount of Zn substitution may be occurred in Zn stearate doped sample [16]. However, further study is needed to clarify this point.

Figure 3 summarizes the field dependence of $J_c$ at 4.2 K for the stearic acid, Mg stearate and Zn stearate doped tapes annealed at 800 °C and 850 °C, respectively. As a reference, the data of the pure tapes which was sintered at 800 ℃ are also included in the figure. Overall, all the doped tapes showed an enhancement of more than one order of magnitude in $J_c$-$H$ performance in fields above 11 T, compared to the pure tapes. The field dependence of $J_c$ was also improved, suggesting that $H_{irr}$ was enhanced by doping, as will be discussed later. Of particular interest is that the $J_c$ of all doped samples annealed at 800 °C are strongly influenced by different dopants. For example, the $J_c$ at 10 T and 4.2 K for Zn stearate and Mg stearate doped tapes exhibited values of $1.84 \times 10^4$ A/cm$^2$ and $1.63 \times 10^4$ A/cm$^2$, respectively, whereas the stearic acid doped tapes showed a lower $J_c$ value of $1.3 \times 10^4$ A/cm$^2$ at the corresponding conditions. We speculated that stearate might have lower decomposition temperatures than that of stearic acid and the C released from stearate can easily substitute for B site, as supported by the different a-axis lattice constants in Table I. In addition, the $J_c$ values of all doped samples were much enhanced when the sintering temperature was further increased to 850 °C. Especially for the stearic acid doped samples, the $J_c$ for this sample exhibits a slightly better field performance and higher values of $J_c$ in high field than other two stearate doped samples. Higher $J_c$ values of $8.6 \times 10^3$ A/cm$^2$ at 12 T and $3.72 \times 10^3$ A/cm$^2$ at 14 T were observed for the stearic acid doped samples sintered at 850 °C. The higher net C percentage of stearic acid than that of other two stearate dopants might be responsible for the better high field performance. However, Zn stearate doped tapes annealed at 850 °C show higher $J_c$ than stearic acid doped tapes in magnetic fields lower than 11 T, and the $J_c$ reached $2.02 \times 10^4$ A/cm$^2$ at 4.2 K and 10 T. This value was comparable to that of optimal SiC doped tapes [17].

To confirm the enhanced flux pinning ability in MgB$_2$ tapes with stearic acid and stearate doping, Figure 4 plots the field dependence of the normalized volume pinning force $F_p$/ $F_P^{max}$ at 20 K for samples heated at 800℃. It is clear that the pinning force for doped tapes is much higher than that of the pure one over 1 T, indicating enhanced flux pinning in high fields. The samples with the Zn stearate doping show the highest flux pinning force among the samples studied: the $F_P^{max}$ of Zn stearate doped samples is up to 1.4 T compared to the 0.7 T of pure sample. As we can see, stearic acid and Mg stearate doped samples also show the shifts of the pinning force curves to higher fields, but the shifts are not as pronounced as for Zn stearate doped sample. Briefly, these results clearly demonstrated that the enhanced flux pinning as a result of stearic acid and stearate doping should be responsible for the excellent transport $J_c$–$H$ properties of doped samples.

The temperature dependence of $H_{c2}$ and $H_{irr}$ for the pure and Zn stearate doped tapes is shown in Figure 5. Clearly, the $H_{c2}$ and $H_{irr}$ of the Zn stearate doped tapes increased more rapidly with decreasing temperature than that of the undoped one. The improved $H_{c2}$ is originated from a reduced coherence length ξ due to the enhanced intraband scattering by C substitution into B sites, as predicted by the theoretical models [10]. The lattice distortion and intragranular inclusions are resulted of the doping. These are believed to increase the intraband scattering, shorten the electron mean free path $l$ and ξ according to the equation $1/\xi=1/l+1/\xi_0$. Actually, higher resistivity of 155.2 μΩcm at 40 K was observed in the Zn stearate doped sample, while 27.4 μΩcm for the pure one. Similar enhancement in $H_{irr}$ can be explained by the enhanced flux pinning as a result of doping. It is also worth noting that at 20 K, the $H_{irr}$ achieved 10 T for Zn stearate doped tapes, which was comparable to the $H_{irr}$ at 4.2 K of NbTi conductors [18]. This result suggests that MRI magnet made by NbTi wires can be replaced with a convenient cryogen-free magnet operated at around 20 K fabricated with $MgB_2$ conductors.

In order to understand the mechanisms for the enhancement of $J_c$-$H$ performance in the doped samples, a TEM study was carried out. Figure 6 shows the typical low-magnification and high-resolution TEM micrographs for the Zn stearate doped tapes. As shown in low-magnification TEM micrographs that the samples are tightly packed $MgB_2$ nanoparticle structure with grain size less than 30nm. Obviously, the fine grain size would create many grain boundaries that may act as effective pinning centres, as in the case of $Nb_3Sn$ superconductor. The high-resolution TEM images also clearly show a large number of dislocations within $MgB_2$ grains, which are similar to those shown in SiC and nanocarbon doped samples [2, 9]. Moreover, the TEM examination revealed that there are a number of impurity phases in the form of nanometer-size inclusions (less than 20 nm in size) inside grains in Zn stearate doped samples. Through EDS analysis, Mg, B, C, O and Zn were detected inside the grains as demonstrated in Fig. 6(c). From XRD data, we believe that full carbon substitution is not achieved, the carbon released from the decomposition of dopants may appear as excess nanoparticles or may react with B or Mg to form BC [19] and $Mg_2C_3$ [9]. These results suggest that the inclusion nanopticles might be unreacted C, BC, $Mg_2C_3$ or MgO and $MgB_4$ detected by XRD. Dislocations and nanosized inclusions are known to serve as strong pinning centers to improve flux pinning [9, 20]. Therefore, the strong flux pinning centers induced by grain boundaries, dislocations, nanosized inclusions as well as carbon substitution effect are responsible for the superior $J_c$-$H$ performance of the doped samples, as shown in Fig.3.

## 4. Conclusions

In summary, we have demonstrated that both the transport $J_c$ and the flux pinning of $MgB_2$ tapes are all significantly enhanced by stearic acid, Mg stearate and zinc stearate doping. The $J_c$ at 12 T and 4.2 K for the zinc stearate doped tape sintered at 850 °C increased by a factor of 21 compared to that of the pure tape. In particular, the zinc stearate tapes showed the $H_{irr}$ value (10T) at 20 K comparable to that of commercial NbTi at 4.2 K. The enhancement of electromagnetic properties in doped samples can be attributed to the lattice distortion resulting from the incorporation of C atoms into the $MgB_2$ crystal lattice as well as a high density of flux-pinning centers obtained in doped tapes. These results suggest that stearic acid and stearate are promising kind of dopants for $MgB_2$ tapes with excellent transport $J_c$ under high field.

**Acknowledgments**

The authors thank S Awaji, G. Nishijima, Ling Xiao, Yulei Jiao, Liye Xiao and Liangzhen Lin for their help and useful discussion. This work is partially supported by the National Science Foundation of China under Grant Nos. 50472063 and 50572104, National '973' Program (Grant No. 2006CB601004) and National '863' Project (Grant No. 2006AA03Z203).

## References

[1] Iwasa Y, Larbalestier D C, Okada M, Penco R, Sumption M D and Xi X 2006 IEEE Trans. Appl. Supercond. 16 1457

[2] Dou S X, Soltanian S, Horvat J, Wang X L, Zhou S H, Ionescu M, Liu H K, Munroe P and Tomsic M 2002 Appl. Phys. Lett. **81** 3419

[3] Sumption M D, Bhatia M, Rindfleisch M, Tomsic M, Soltanian S, Dou S X and Collings E W 2005 Appl. Phys. Lett. **86** 092507

[4] Bugoslavsky Y, Cohen L F, Perkins G K, Polichetti M, Tate T J, Gwilliam R and Caplin A D 2001 Nature **410** 561

[5] Senkowicz B J, Giencke J E, Patnaik S, Eom C B, Hellstrom E E and Larbalestier D C 2005 Appl. Phys. Lett. **86** 202502

[6] Ma Y, Kumakura H, Matsumoto A, Hatakeyama H and Togano K 2003 Supercond. Sci. Technol. **16** 852

[7] Lezza P, Senatore C and Flükiger R, 2006 Supercond. Sci. Technol. **19**, 1030

[8] Ma Y, Zhang X, Nishijima G, Watanabe K, Awaji S and Bai X 2006 Appl. Phys. Lett. **88** 072502

[9] Yamada H, Hirakawa M, Kumakura H, and Kitaguchi H, 2006 Supercond. Sci. Technol. **19**, 175

[10] Gurevich A, 2003 Phys. Rev. B **67** 184515

[11] Gurevich A *et al* 2004 Supercond. Sci. Technol. **17** 278

[12] Kim J H, Zhou S, Hossain M S A, Pan A V and Dou S X 2006 Appl. Phys. Lett. **89** 142505

[13] Lee S, Masui T, Yamamoto A, Uchitama H, and Takama S, 2003 Physica C **397**, 7

[14] Yamamoto A, Shimoyama J, Ueda S, Katsura Y, Iwayama I, Horii S, and Kishio K, 2005 Appl. Phys. Lett. **86**, 212502

[15] Wilke R H T, Bud'ko S L, Canfield P C, Finnemore D K, Suplinskas R J, Hannahs S T, 2004 Phys. Rev. Lett. **92**, 217003

[16] Kazakov S M, Angst M, Karpinski J, Fita I M and Puzniak R, 2001 Solid State Commun. **119,** 1

[17] Matsumoto A, Kumakura H, Kitaguchi H, and Hatakeyama H, 2004 Supercond.Sci.Technol.

**17,** S319

[18] Larbalestier D, Gurevich A, Feldmann D M, and Polyanskii A, 2001 Nature (London) **414**, 368

[19] Dou S X, Braccini V, Soltanian S, Klie R, Zhu Y, Li S, Wang X L and Larbalestier D 2004 J. Appl. Phys. **96** 7549

[20] Serquis A, Civale L, Hammon D L, Liao X Z, Coulter J Y, Zhu Y T, Jaime M, Peterson D E, Mueller F M, Nesterenko V F and Gu Y, 2003 Appl. Phys. Lett. **82**, 2847

TABLE I. Measured data for pure and doped MgB$_2$ tape samples heated at 800 ℃.

| Sample | Lattice | | Actual C (x) in Mg(B$_{1-x}$C$_x$)$_2$ | $T_c$ (K) | RRR | $\rho_{40}$ (μΩcm) | FWHM of (*110*) peak (º) | $J_c$ (4.2 K) | |
|---|---|---|---|---|---|---|---|---|---|
| | $a$ (Å) | $c$ (Å) | | | | | | 10 T | 14 T |
| Pure | 3.0844 | 3.5265 | | 37.9 | 2.147 | 27.4 | 0.319 | 2.1×10$^3$ | |
| Zn stearate | 3.0758 | 3.5283 | 0.0194 | 35.0 | 1.560 | 155.2 | 0.586 | 1.8×10$^4$ | 2.9×10$^3$ |
| Mg stearate | 3.0756 | 3.5274 | 0.0198 | 35.6 | 1.575 | 112.8 | 0.588 | 1.6×10$^4$ | 2.4×10$^3$ |
| Stearic acid | 3.0771 | 2.5311 | 0.0164 | 35.5 | 1.552 | 126.7 | 0.598 | 1.3×10$^4$ | 2.2×10$^3$ |

# Captions

Figure 1  XRD patterns of in situ processed pure and doped tapes sintered at 800 ℃. The peaks of $MgB_2$ indexed, while the peaks of MgO and $MgB_4$ are marked by * and #, respectively.

Figure 2  Normalized magnetic susceptibility versus temperature for pure and doped tapes heated at 800 ℃. The inset shows an enlarged view near the superconducting transitions.

Figure 3  $J_c$-$H$ properties of Fe-sheathed pure and doped tapes. The measurements were performed in magnetic fields parallel to the tapes surface at 4.2 K.

Figure 4  Normalized volume pinning force ($F_P/F_P^{max}$) versus magnetic field (T) at 20 K for pure and doped tapes heated at 800 ℃.

Figure 5  Temperature dependence of $H_{irr}$ and $H_{c2}$ for the pure and Zn stearate doped tapes

Figure 6  TEM images showing the intragrain dislocations and nanoparticle inclusions of the the Zn stearate doped samples sintered at 800 ℃. (a) Low magnification micrograph for Zn stearate doped samples. (b) High magnification micrograph for Zn stearate doped samples. (c) EDS element analysis of $MgB_2$ grains, a small Cu-peak appeared due to the background signal from the sample holder.

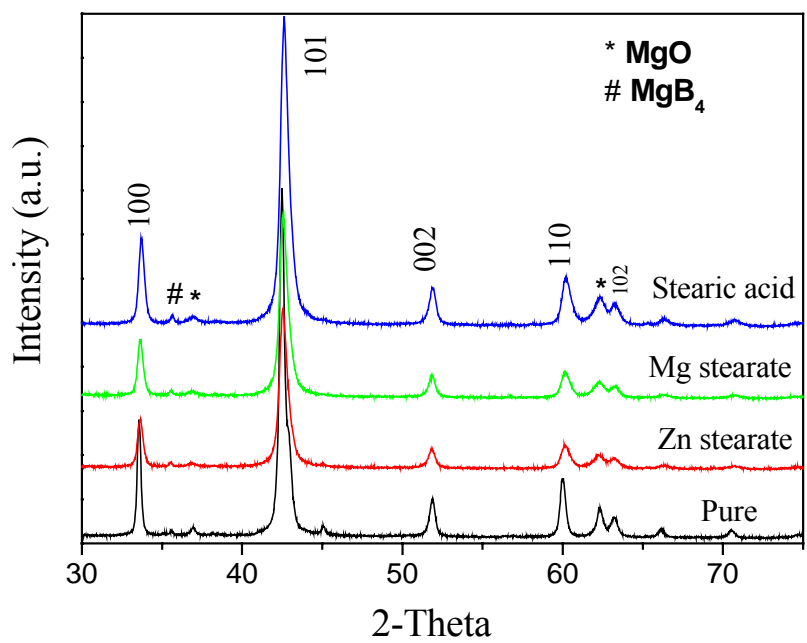

Fig.1 Gao et al.

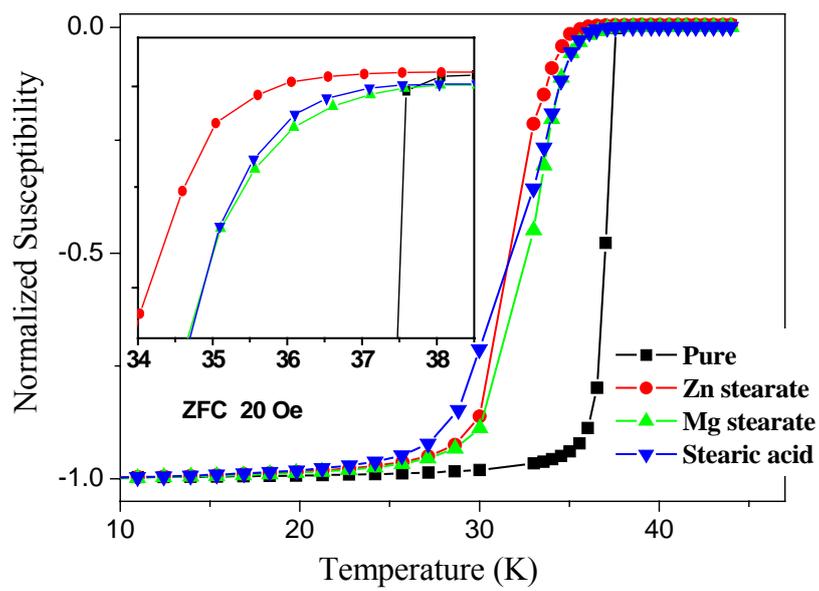

Fig.2 Gao et al.

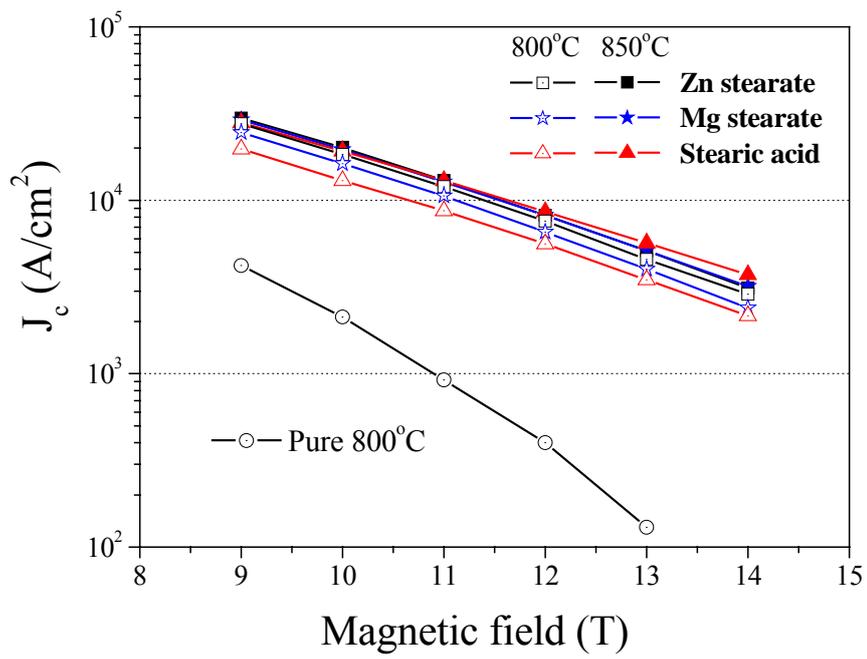

Fig.3 Gao et al.

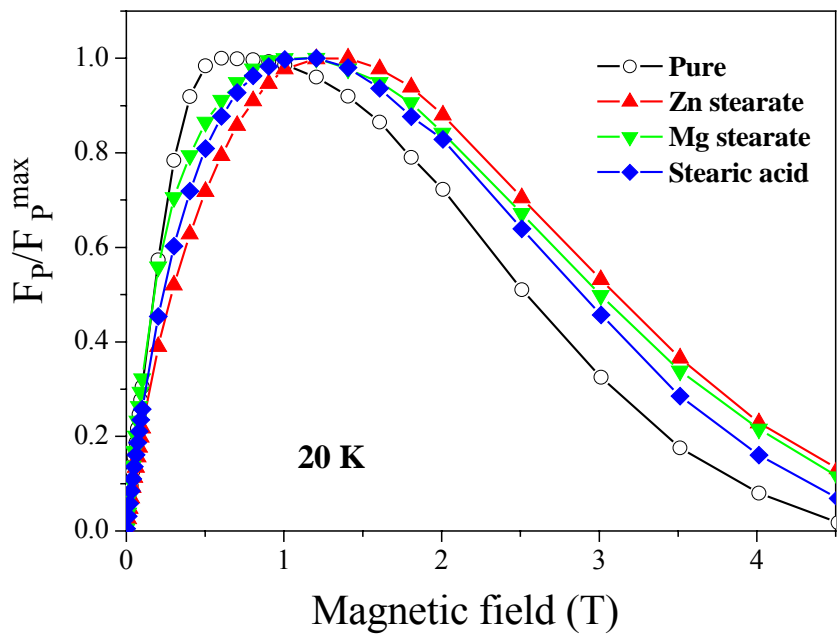

Fig.4 Gao et al.

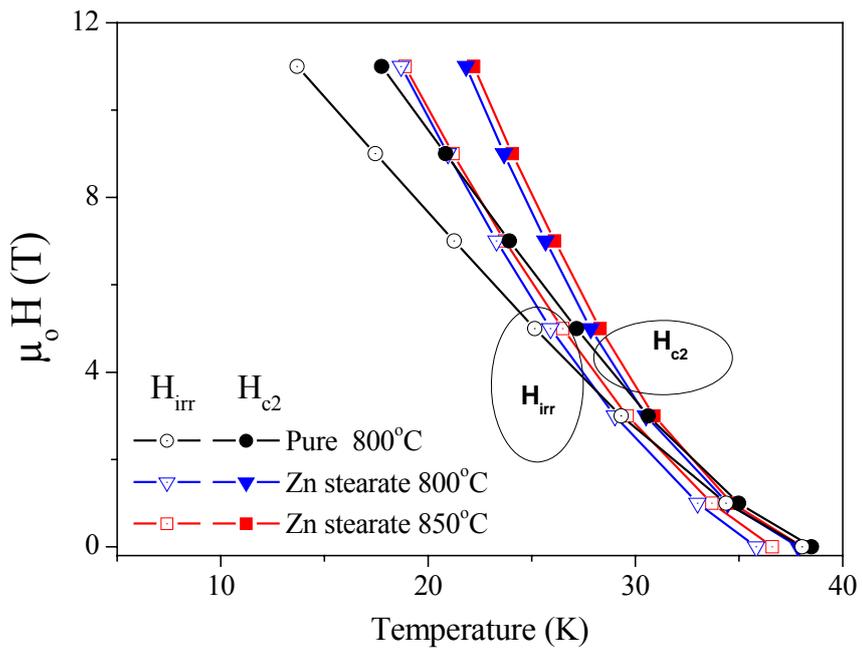

Fig.5 Gao et al.

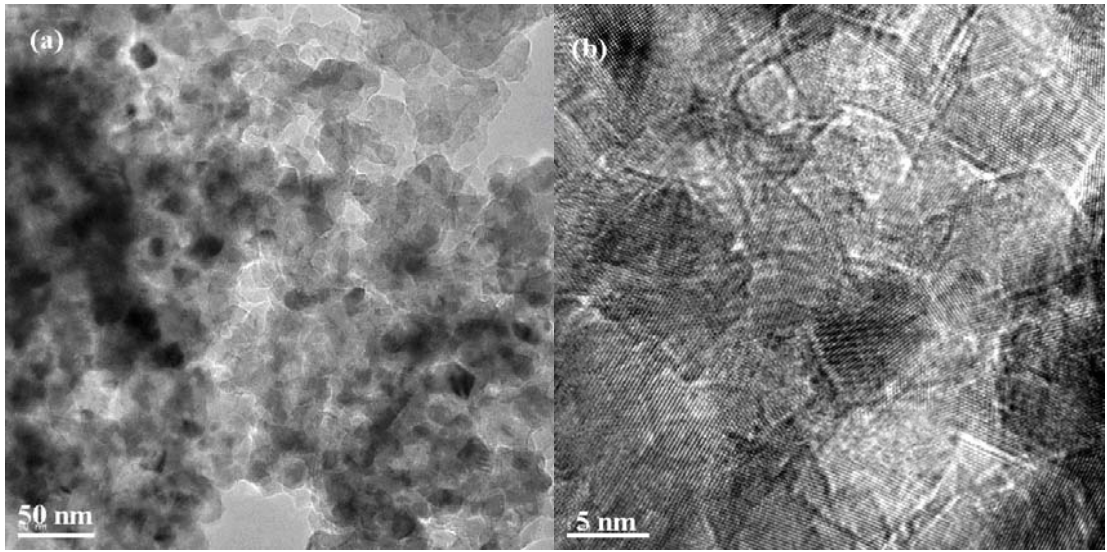

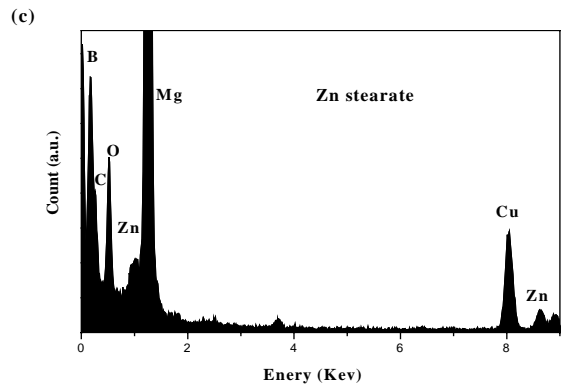

Fig.6 Gao et al.